\documentclass[prb,twocolumn,showpacs,floatfix]{revtex4}
\usepackage{graphicx}
\usepackage{dcolumn}
\usepackage{epstopdf} 
\usepackage{bm}
\usepackage{color} 
\newcommand{\sstrut}{\rule[-1ex]{0ex}{3ex}} 
\newcommand{\fig}[1]{figure~\ref{#1}}
 \newlength{\figwidth}
\newcommand{\tabl}[1]{table~\ref{#1}}
 
\newlength{\hw} 
\newlength{\minusspace} 
\newcommand{\msp}{\hspace{\minusspace}} \newlength{\zerospace}
 \newcommand{\zsp}{\hspace{\zerospace}}
\newenvironment{tablenotes}{\begin{list}{$^{\alph{enumi}}$}
{\usecounter{enumi}\setlength{\itemsep}{0ex} \setlength{\topsep}{-2ex}}}
{\end{list}} 
\newcommand{\mub}{$\mu_{\mathrm B}$}
\newcommand{\sinth}{\ensuremath{\sin\theta/\lambda}}
\newcommand{\inA}{\AA$^{-1}$} 
 
\newcommand{\baf}{BaFe$_2$As$_2$} \newcommand{\kv}{\ensuremath{{\bf k}}}
\newcommand{\ku}{\ensuremath{{\bf{\hat k}}}}

\begin{document} 
\title{\bf \large Magnetisation distribution in
the tetragonal phase of \baf} 
\author{P. J. Brown$^{1,2}$, T.
Chatterji$^3$,  A. Stunault$^1$, Y. Su$^4$, Y. Xiao$^5$, R. Mittal$^{4,6}$,
Th. Br\"uckel$^{4,5}$, Th. Wolf$^7$, and P. Adelmann$^7$ } 
\affiliation{$^1$
Institut Laue-Langevin, B.P. 156, 38042 Grenoble Cedex 9, France\\
$^2$Department of Physics, Loughborough University, LE11 3TU, UK\\
$^3$JCNS, Forschungszentrum J\"ulich, Outstation at Institut
Laue-Langevin, B.P. 156, 38042 Grenoble Cedex 9, France\\ $^4$ JCNS,
Forschungszentrum J\"ulich, Outstation at FRMII, Lichtenbergstrasse 1,
85747 Garching, Germany\\ $^5$ Institut f\"ur Fesk\"orperforschung,
Forschungzentrum J\"ulich, 52425 J\"ulich, Germany\\
$^6$Solid State Physics Division, Bhabha Atomic Research Centre, Trombay, Mumbai 400085, India\\
$^7$Karlsruhe Institute of Technology, Institut fuer Festkoerperphysik, D-76021 Karlsruhe, Germany} 
\date{\today}

\begin{abstract} 
We have determined the spatial distribution of the
magnetisation induced by a field  of 9~T in the tetragonal phase of
\baf\ using polarised neutron diffraction. Magnetic strcture factors
derived from the polarisation dependence of the intensities of Bragg 
reflections  were used to make a maximum entropy reconstruction of the
distribution projected on the 110 plane. The reconstruction shows
clearly that the magnetisation is confined to the region around the iron
atoms and that there is no significant magnetisation associated with
either the As or Ba atoms. The distribution of magnetisation around the
Fe atom is significantly non-spherical with a shape which is extended in
the $\langle111\rangle$ directions in the projection. These results show
that  the electrons which give rise to the paramagnetic susceptibility
are confined to the Fe atoms their distribution suggests that they
occupy $3d$ $t_{2g}$ type orbitals with $\approx60$\% in those of $xy$
symmetry. 
\end{abstract} 
\maketitle The pnictide
superconductors and their parent compounds have drawn extensive
attention because they provide a new opportunity to investigate the
mechanism of non-BCS exotic superconductivity
\cite{Kamihara,Takahashi,Chen1,Matsuishi1,Rotter1}. Most of the research
on pnictide superconductors has focused on two classes of compounds,
\emph{R}FeAs(O$_x$F$_{1-x}$)(with \emph{R} = La, Nd and Sm) and
\emph{A}Fe$_2$As$_2$ (with \emph{A} = Ba, Ca and Sr), the so called
'1111' and '112' families. Both these two classes of compounds adopt a
layered structure with a single FeAs layer in the unit cell of '1111'
and two such layers in the unit cell of '122'. The superconducting state
can be induced either by electron or hole doping of the parent compounds
or also by pressure \cite{Wen,Ren,Matsuishi2}. Till now, the highest
\emph{T}$_c$ attained is 57.4 K in the electron doped
Ca$_{0.4}$Na$_{0.6}$FeAsF '1111' compound \cite{Cheng}, while for '122'
family the highest \emph{T}$_c$ of 38 K is reached in the hole doped
Ba$_{0.6}$K$_{0.4}$Fe$_2$As$_2$ \cite{Rotter2}. The '122' compounds 
differ from the cuprate superconductors in that the superconducting state can be induced
by the application of pressure only \cite{mcqueeney08,alireza09}. It seems that the
FeAs layers are responsible for superconductivity in these compounds
because the electronic states near the Fermi surface are dominated by
contributions from Fe and As.

Recent neutron diffraction experiments reveal that the common feature of
all the iron pnictide parent compounds
\cite{Cruz,Huang,Su1,Goldman,Xiao} is a spin density wave (SDW) arising
from long range antiferromagnetic (AFM) order of the Fe moments at low
temperature. For the parent compounds the onset of AFM order coincides
with the tetragonal-orthorhombic (T-O) structural phase transition in
the '122' family and is preceded by it in the '1111' family. The role of
orbital ordering in driving these transitions and leading to anisotropic
magnetic coupling is still being debated \cite{Lee_09}. Phase diagrams
of some iron pnictides show clearly that the magnetic order can be
suppressed by charge carrier doping of the parent compound.
Concomitantly, superconductivity emerges and reaches a maximum
\emph{T}$_c$ at optimal doping~\cite{Zhao1}, thus exhibiting features
similar to high \emph{T}$_c$ cuprates \cite{Bednorz}. Extensive studies
of phonon dynamics \cite{Ranjan1,Ranjan2} suggest that it is unlikely 
that the superconductivity in iron pnictides is due to simple
electron-phonon coupling. Since it seems that phonons play no
significant role in the superconducting pair formation, it is natural to
presume that  magnetism has a crucial role in the appearance of
superconductivity and consequently AFM spin fluctuations have been
suggested as a possible pairing mechanism. Strong evidence for the
presence of resonant spin excitations in the superconducting phase has
indeed been obtained from recent inelastic neutron scattering
experiments on several optimally doped '122' superconductors
\cite{Christianson,Lumsden,Chi}.

The nature of magnetism and possible orbital order in iron pnictide
compounds are still very controversial and therefore additional
experimental information on these degrees of freedom for the parent
compounds can be helpful in understanding the nature of
superconductivity in these compounds. In order to get direct information
about the electronic structure of the parent '122' compound we have undertaken
a polarised neutron diffraction experiment on BaFe$_2$As$_2$ to
determine the field induced magnetisation distribution. A good quality
single crystal was grown by the self-flux method. The structural
parameters were determined from unpolarised neutron integrated intensity
measurements made using the 4-circle diffractometer D9 and flipping
ratios were measured using the polarised neutron diffractometer D3. Both
these instruments are installed on the hot neutron source of the high-flux
reactor of the Institute Laue-Langevin in Grenoble. The sample was held
at constant temperature in a closed-cycle refrigerator on D9 whereas on
D3 it was oriented with a $\langle1\bar{1}0\rangle$ axis parallel to the
vertical field direction of a 9 Tesla cryomagnet. The flipping ratios
from the crystal were measured in the paramagnetic tetragonal phase at
$T = 200$ K.

Sets of experimental structure factors containing 70 independent
reflections  $\sinth<0.85$~\inA measured with $\lambda=0.84$~\AA\ and 90
with $\sinth<1.0$~\inA and $\lambda=0.52$~\AA\ were obtained from the
integrated intensities measured on D9 after averaging the intensities
over equivalent reflections. These data were used in least squares
refinements of the crystal structure in which the variable parameters
were the $z$ coordinate of As, the isotropic temperature factors for the
three sites  and a single 
\begin{table}[htdp]{\footnotesize
\vspace{-1ex} \caption{Parameters obtained in least squares refinements
of integrated intensities measured at T = 200 K on D9.} \begin{center}
\begin{tabular}{lccllll} \hline
Atom&\multicolumn{2}{c}{Position in I4/mcm} &\multicolumn{1}{c}{$z$}&B (\AA$^2$)\\ \hline
Ba&2a&$0\ 0\ 0$&&0.62(4)\\
Fe&4d&$\frac12\ 0\ \frac14$&&0.43(3)\\
As&4e&$0\ 0\ z$&0.3543(1)&0.55(3)\\ \hline
\multicolumn{2}{l}{Extinction}& g
(rad$^{-1}$)&\multicolumn{1}{c}{1.4(1.6)}\\ $R_{\mbox{\scriptsize
cryst}}$&$\lambda=0.84$ \AA&3.8&$\lambda=0.51$ \AA&2.7\\ 
\hline \end{tabular} \end{center}
\label{strucpars} \vspace{-2ex} }\end{table}
extinction parameter $g$
representing the mosaic spread of the crystal. The results are
summarised in \tabl{strucpars}. The small value obtained for $g$, which
is less than its estimated error, shows that any extinction is very
small.

The ratios between the intensity scattered by the Bragg reflections  in the $[1\bar10]$
zone for incident neutrons polarised
parallel and anti-parallel to the applied field of 9~T  (polarised neutron flipping ratios) were
measured  at 200~K using a neutron wavelength 0.825 \AA. 
Since the susceptibility of \baf\ is small $ < 5 \times 10^{-4}$ emu/mole, all the
flipping ratios $R$ are close to unity and since the magnetic structure
factors are proportional to $R$-1, it was necessary to record more than
$10^7$ neutrons from each reflection to obtain $\approx5$\% precision.
The  flipping ratios measured for equivalent reflections and for
repeated measurements of the same reflection were averaged together to
give a mean value of $R$ and used to calculate the magnetic structure
factors $F_M$ using the relationship \[F_M=\frac{(R-1)F_N}{2(
P^++P^-)}\] where $P^+$ and $P^-$ are the efficiencies of neutron
polarisation parallel and antiparallel to the applied field; $F_N$ is
the nuclear structure factor which was calculated using the parameters
obtained from the integrated intensity measurements 
which are given in  table~\ref{strucpars}. 

The magnetisation induced in a crystal from the same 
batch as the one used in the experiment, by a field of  9~T applied
the 001 plane at 200~K was measured as 0.0100 \mub/f.u.. It is the sum 
of a paramagnetic part due to magnetic excitation of electrons near the 
Fermi surface and a diamagnetic part to which all electrons contribute. 
The diamagnetic volume susceptibility is given by the Langevin equation.
\[\chi_{dia}=-(e^2/6Vmc^2)\sum_i Z_i\langle r^2\rangle_i\]
the sum is over all the atoms $i$ in the unit cell of volume V, $\langle r^2\rangle_i$ 
is the mean square radius of the $i$th atom's electron
wave function and $Z_i$ its atomic number. The diamagnetic contribution
to the magnetic structure factor is
\begin{equation}F_{dia}=\frac{HC}{|\kv|}\sum_i\frac{df(k)_i}{dk}
\exp{\imath\kv\cdot{\bf r}_i}\label{fdia}\end{equation}
where $f_i(k)$ is the atomic form factor of the $i$th atom and ${\bf r}_i$
its position in the unit cell. The constant C has the value 
$1.52 \times 10^{-5}$~\mub$T^{-1}$\AA$^2\quad$\cite{Stassis_70,Maglic_78}.
The diamagnetic contribution to the magnetisation calculated using the atomic form 
factors for Ba, Fe and As \cite{IntTab} is -0.0033\mub/f.u., the paramagnetic 
part of the magnetisation is therefore  $0.0100 -(-0.0033)=0.0133$~\mub/f.u.. 
The diamagnetic 
contributions to the magnetic structure factors were calculated using eqn.~\ref{fdia}
and are given in \tabl{strucfacs}.
The values $F_{dia}$ were subtracted from the magnetic structure factors $F_M$ obtained 
from the flipping ratios to give the paramagnetic structure factors $F_{para}$
also listed in \tabl{strucfacs}.
\begin{table}[hbt]\vspace{-1ex}{\footnotesize
\caption{Observed and calculated magnetic structure factors for the
tetragonal phase of \baf at 200~K.} 
\begin{center}
\begin{tabular}{lllcllll} 
\hline 
&&&\sinth&$\quad F_{dia}^a$&\quad $F_{para}^b$&\quad $F_{calc}^c$&\quad $F_{calc}^d$\\[-1.5ex]
$h$&$k$&$l$\\[-1.5ex]
&&&{\scriptsize(\inA)}&\quad {\scriptsize(m\mub)}&\quad {\scriptsize(m\mub)}&\quad {\scriptsize(m\mub)}
&\quad {\scriptsize(m\mub)}\\
\hline 
$  0$&$  0$&$  2$&$   0.077$&$\msp  0.48$&$  -23.7(1.1)$&$   -25.5$&$   -24.4$\\
$  0$&$  0$&$  4$&$   0.154$&$ -0.77$&$\msp   19.6(1.1)$&$\msp    21.3$&$\msp    20.9$\\
$  1$&$  1$&$  2$&$   0.195$&$ -1.09$&$\msp   18.3(1.4)$&$\msp    18.6$&$\msp    17.5$\\
$  1$&$  1$&$  4$&$   0.236$&$\msp  0.59$&$      -17(2)$&$   -15.8$&$   -15.4$\\
$  1$&$  1$&$  6$&$   0.292$&$ -1.06$&$\msp   12.5(1.4)$&$\msp    12.2$&$\msp    12.5$\\
$  0$&$  0$&$  8$&$   0.308$&$ -0.92$&$\msp       13(3)$&$\msp    11.3$&$\msp    11.9$\\
$  2$&$  2$&$  0$&$   0.357$&$ -0.88$&$\zsp\msp    8.0(1.4)$&$\zsp\msp     8.7$&$\zsp\msp     8.1$\\
$  2$&$  2$&$  2$&$   0.366$&$\msp  0.25$&$      -12(2)$&$\zsp    -8.3$&$\zsp    -7.9$\\
$  0$&$  0$&$ 10$&$   0.386$&$\msp  0.41$&$\zsp      -10(2)$&$\zsp    -7.4$&$\zsp    -8.1$\\
$  2$&$  2$&$  4$&$   0.389$&$ -0.32$&$\zsp\msp       10(2)$&$\zsp\msp     7.3$&$\zsp\msp     7.3$\\
$  1$&$  1$&$ 10$&$   0.425$&$ -0.24$&$\zsp\msp        6(2)$&$\zsp\msp     5.9$&$\zsp\msp     6.7$\\
$  0$&$  0$&$ 12$&$   0.463$&$ -0.38$&$\zsp\msp        7(4)$&$\zsp\msp     4.6$&$\zsp\msp     5.2$\\
$  2$&$  2$&$  8$&$   0.472$&$ -0.45$&$\msp       10(3)$&$\zsp\msp     4.4$&$\zsp\msp     5.3$\\
$  1$&$  1$&$ 12$&$   0.496$&$\msp  0.10$&$\zsp       -2(5)$&$\zsp    -3.7$&$\zsp    -4.5$\\
$  2$&$  2$&$ 10$&$   0.526$&$\msp  0.23$&$\zsp       -4(4)$&$\zsp    -3.0$&$\zsp    -4.1$\\
$  3$&$  3$&$  2$&$   0.542$&$ -0.22$&$\zsp       -2(5)$&$\zsp\msp     2.7$&$\zsp\msp     3.2$\\
\hline 
\end{tabular}
\end{center} 
\begin{tablenotes} 
\item Diamagnetic contribution to the magnetic structure factor
\item$F_{para}=F_M-F_{dia}$
\item Structure factors calculated with
an Fe moment of 0.0068 \mub and a spherically symmetric neutral Fe form
factor \cite{freemanW} 
\item{Structure factors calculated using the
multipole model with the parameters of the $t_{2g}$ only model in
\tabl{mpole}} 
\end{tablenotes} 
\label{strucfacs} }
\end{table}

An effective form factor for the Fe atom, 
\begin{figure}[hbt]\vspace{-1ex} \begin{center}
\resizebox{0.45\textwidth}{!}{\includegraphics{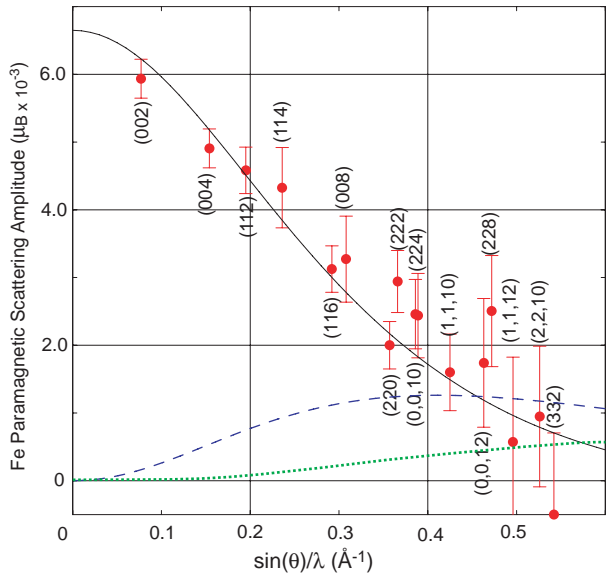}}
\caption{Paramagnetic scattering amplitudes 
measured for Fe in \baf\ at
200~K. The solid curve shows the $\langle j_0\rangle$ form
factor for neutral Fe$\quad$\cite{freemanW}, scaled to the paramagnetic magnetisation
of $6.65\times 10^{-3}$\mub/Fe. The dashed (blue) and dotted (green) curves show the
$\langle j_2\rangle$ and $\langle j_4\rangle$	 form factors, which multiply the anisotropic terms
in the magnetic scattering on the same scale} \label{formfac}
\end{center}\vspace{-5ex} 
\end{figure} 
obtained by dividing each $F_{para}$ by the geometric structure factor of
Fe for that reflection (4 for $h+k+l$ even and -4 for
$h+k+l$ odd) is shown in \fig{formfac} where it is compared 
with the Fe 3d free atom curve \cite{freemanW} scaled to 6.65~\mub. 
The low angle reflections fall on the curve within experimental error
but at higher angles, at which the higher order form factors 
$\langle j_2\rangle$ and $\langle j_4\rangle$	 become appreciable,
significant scatter is apparent which may characterise an aspherical
magnetisation distribution. 

\begin{figure}[hbt]
\begin{center}
\hfil\parbox{0.35\textwidth}{\resizebox{0.3\textwidth}{!}
{\includegraphics{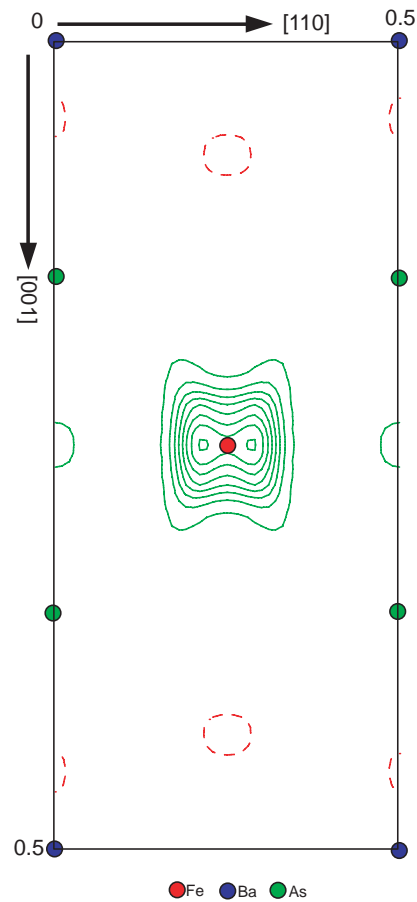}}}
\parbox{0.45\textwidth}{ \caption{Maximum
entropy reconstruction of the magnetisation distribution in tetragonal \baf\ at
200~K projected down $[1\bar10]$. Contours are drawn at intervals of
$10^{-2}$ \mub\AA$^{-2}$.} \label{maxemap}}\hfil
\end{center}\vspace{-3ex} \end{figure}

The method of maximum entropy \cite{skilling_89, papoular_90}
provides a model free method for reconstructing  an image from sparse and noisy
data. We have used this method to clarify  the shape of the distribution. 
 The maximisation procedure coded in the MEMSYS III subroutine library 
\cite{memsys3} was used  to make the maximum entropy
reconstruction of the magnetisation distribution projected down
$[1\bar10]$, from the measured magnetic structure factors. The result of 
the reconstruction is shown in \fig{maxemap}.
The reconstruction shows clearly
that the magnetisation is confined to the region around the iron atoms
and that there is no significant magnetisation associated with either
the As or Ba atoms. The magnetisation around the Fe atom is
significantly non-spherical with a shape that appears to extend in the
[111] directions of the projection. Further clarification of the shape
of the iron atom magnetisation was obtained by fitting the magnetic
structure factors to a multipole model in which they are expressed as
\[F_M(\kv)=a_0\langle j_0(|\kv|)\rangle + \sum_{l=2,4}\langle
j_l(|\kv|)\rangle\sum_{m=-l}^{m=l}a_{lm} Y_{\ku}(lm\pm)\] where the $\langle
j_l(|\kv|)\rangle$ are the form factor integrals for a neutral Fe atom
\cite{freemanW} and the  $Y_{\ku}(lm\pm)$ are the real combinations of spherical
harmonic functions: 
\[ Y_{\ku}(lm\pm) =
\frac1{\sqrt2}\left(Y_l^{-m}(\ku) \pm(-1)^m Y_l^{m}(\ku)\right) 
\]%
\begin{table}[thb]{\footnotesize \setlength{\hw}{1.5ex} \caption{Fe
Multipole amplitudes and 3d orbital occupancies determined from the
magnetic structure factors determined for \baf\  at 200~K.}
\begin{center} \parbox{0.45\textwidth}{ \begin{tabular}{lcrr} \hline
&&\multicolumn{2}{c}{Amplitudes (m\mub)}\\[-\hw] Function&Coeff.\\[-\hw]
&&All $d^a$&$t2_{g}$ only$^b$\\ \hline
$Y(00)$&$a_0$&$\msp6.3(2)$&$\msp6.4(2)$\\
$Y(20)$&$a_{20}$&$-1.6(9)$&$-1.15(6)$\\
$Y(40)$&$a_{40}$&$-2(3)$&$-1.3(4)$\\
\sstrut $Y(44+)$&$a_{44}$&$-8(4)$&$-2.8(5)$\\
\hline 
$\chi^2$&&$0.87$&$0.88$\\ 
\hline 
\end{tabular} }
\hfill\parbox{0.45\textwidth}{ \begin{tabular}{crr} \hline
&\multicolumn{2}{c}{Occupancies (\%)}\\[-\hw] Orbital\\[-\hw] &All
$d^a$&$t2_{g}$ only$^b$\\ \hline $3z^2-r^2$&$ -16(28)$&$ 0$\\
$x^2-y^2$&$-42(36)$&$ 0$\\ $xy$&$\msp 98(36)$&$52(6)$\\ $xz$, $yz$&$\msp
61(6)$&$48(6)$\\ \hline \end{tabular} } \end{center} \begin{tablenotes}
\item All multipole parameters allowed by $\bar4m2$ point symmetry. \item
Multipole parameters constrained to give only $t_{2g}$ type orbitals.
\end{tablenotes} \label{mpole}\vspace{-1ex} }\end{table}
The point
group symmetry of the Fe site, $\bar4m2$, limits the non-zero coefficients
$a_{lm}$ to $a_{20}$, $a_{40}$ and $a_{44}$, and the values of the four
coefficients obtained from the least squares fit are given in
\tabl{mpole}. 
 
In a site
with 4-fold symmetry  the $d$ electron orbitals split into three singlet
states: $d_{3z^2-r^2}$, $d_{x^2-y^2}$, $d_{xy}$ and a doublet
combination of $d_{xz}$ and $d_{yz}$. The first two singlet states are
derived from the cubic $e_g$ functions and the third singlet and the
doublet from the $t_{2g}$ ones. The occupancies of these four
non-degenerate orbitals can be derived directly from the coefficients
$a_{lm}$. However the parameters obtained from the unconstrained fit
lead to unphysical, negative occupancies for the two  $e_g$ type
orbitals but with large estimated standard deviations. A constrained fit
in which the ratio between the $a_{lm}$ were fixed to correspond to
occupancy of the $t_{2g}$ type orbitals only, gave  equally good 
 agreement as shown in \tabl{mpole}. The magnetic
structure factors calculated for this constrained multipole model and
also those obtained for the best spherically symmetric model are given
together with the measured values and the diamagnetic corrections in \tabl{strucfacs}.

The magnetic form factor of Fe in the antiferromagnetic phase of the
closely related pnictide superconductor SrFe$_2$As$_2$ has been studied
in two recent investigations\cite{Ratcliff,Lee}. Whereas one of the
publications \cite{Ratcliff} concludes that the magnetisation
distribution is significantly extended in the directions of the FeAs
bonds, the DFT calculations made in the other \cite{Lee}, which also
predicts significant anisotropy in the magnetisation distribution around
the Fe atom, suggests that the most significant extension is rather in
the $\langle100\rangle$ and $\langle110\rangle$ directions.  The
apparently large anisotropy reported by Ratcliff et al.\cite{Ratcliff}
was deduced from  Fourier inversion of the antiferromagnetic form
factor. It is probably largely an artefact introduced because the
experiment only measures the Fourier components of magnetisation in the
plane perpendicular to the magnetisation direction and hence the Fourier
inversion lacks components which would modulate the density in
directions perpendicular to the spin. The apparent extension is
accentuated by the intervention of nodal planes characterising the
antiferromagnetic arrangement. The paramagnetic magnetisation
distribution measured in the present experiment is projected on the
plane perpendicular to the magnetisation direction so that Fourier
components with all orientations in the plane of projection can be
measured. It has the same periodicity as the crystal lattice and so
cannot be compared directly with an antiferromagnetic  magnetisation
distribution which has systematic nodes imposed by the antiferromagnetic
structure.  

The polarised neutron technique has been widely used to determine the distribution
of electrons giving rise to the paramagnetism in many systems. The classical work
on paramagnetic metals is reviewed by Moon \cite{moon_86}, and its application to
cuprate superconductors by Boucherle et al.\cite{bouch_93}. In all cases
the paramagnetic magnetisation arises from redistribution,
by the magnetising field,  of electrons of opposite spin in states near
the Fermi surface and the magnitude of their contribution is proportional 
to their density of states at the Fermi surface. These electrons will only be the 
same as those giving rise to the antiferromagnetic moment if that moment
is due to unpaired states in narrow bands just below the Fermi surface.

The results of the present experiment show that at least 96\% of the electrons 
in \baf\ which give rise to the paramagnetic
susceptibility, are localised on the Fe atoms with a radial distribution
similar to that of a neutral Fe atom. Their angular distribution shows
that they occupy the $t_{2g}$ type orbitals with a strong  preference
for the singly degenerate $xy$ type which has its maxima in the
$\langle110\rangle$ directions which are not those of any ligand atoms
rather than the doubly degenerate $xz$ and $yz$ types which  maximise in
a cone containing  directions nearly parallel to the Fe-As bond
directions. This anisotropy is broadly in agreement with the results of
the DFT calculations \cite{Lee} for antiferromagnetic
SrFe$_2$As$_2$. If, as might be expected, there is strong hybidisation
between the Fe and As atoms these hybridised bonding and antibonding states
must lie well below and well above the Fermi level leaving narrow 3d non-bonding
bands at the Fermi surface.

\end{document}